\documentclass[12pt]{article}

\pdfoutput=1

\usepackage[top=80pt,bottom=85pt,left=85pt,right=85pt]{geometry}
\usepackage{amssymb}
\usepackage{amsmath}
\usepackage{graphicx,color,subfigure}
\usepackage{setspace}
\usepackage{sectsty}
\usepackage[debug,pageanchor=false]{hyperref}
\definecolor{link}{rgb}{.8,.15,.1}
\hypersetup{colorlinks=true,linkcolor=link,citecolor=link,urlcolor=link,linktocpage}
\usepackage{cite}
\usepackage{cancel}
\usepackage{braket}

\newcommand{\R}{\mathbb{R}}
\newcommand{\C}{\mathbb{C}}
\newcommand{\zz}{\mathbb{Z}}

\newcommand{\vol}{\mathrm{vol}}
\newcommand{\A}{\mathcal{A}}
\newcommand{\Athree}{\mathcal{A}_3}
\newcommand{\Ftwo}{\mathcal{F}_2}
\newcommand{\Ffour}{\mathcal{F}_4}
\newcommand{\AdS}{\mathrm{AdS}}

\makeatletter
\@addtoreset{equation}{section}
\makeatother

\begin{document}

\begin{titlepage}

\begin{center}

\vskip .3in \noindent

{\Large \bf{Universal consistent truncation \\ \vspace{.2cm} for 6d/7d gauge/gravity duals}}

\bigskip

Achilleas Passias, Andrea Rota and Alessandro Tomasiello\\

\bigskip
{\small 
Dipartimento di Fisica, Universit\`a di Milano--Bicocca, Piazza della Scienza 3, I-20126 Milano, Italy \\ and \\ INFN, sezione di Milano--Bicocca
}

\vskip .3cm
{\small \tt achilleas.passias, andrea.rota, alessandro.tomasiello @unimib.it}
\vskip .6cm
     	{\bf Abstract }
\vskip .1in
\end{center}

Recently, AdS$_7$ solutions of IIA supergravity have been classified; there are infinitely many of them, whose expression is known analytically, and with internal space of $S^3$ topology. Their field theory duals are six-dimensional $(1,0)$ SCFT's. In this paper we show that for each of these AdS$_7$ solutions there exists a consistent truncation from massive IIA supergravity to minimal gauged supergravity in seven dimensions. This theory has an SU(2) gauge group, and a single scalar, whose value is related to a certain distortion of the internal $S^3$. This explains the universality observed in recent work on AdS$_5$ and AdS$_4$ solutions dual to compactifications of the $(1,0)$ SCFT$_6$'s. Thanks to previous work on the minimal gauged supergravity, the truncation also implies the existence of holographic RG-flows connecting those solutions to the AdS$_7$ vacuum, as well as new classes of IIA AdS$_3$ solutions. 

\noindent

\vfill
\eject

\end{titlepage}

\tableofcontents

\section{Introduction} 
\label{sec:intro}

Conformal field theories in dimensions higher than four are still comparatively mysterious; there is usually no Lagrangian description. This is the case for example for the $(2,0)$-supersymmetric theory living on the world-sheet of coincident M5-branes. Some indirect information can be obtained by compactifying the theory. Reducing it on a $T^2$ gives ${\cal N}=4$ super-Yang--Mills. Reducing it on a Riemann surface produces a vast ``class S'' of four-dimensional theories with very interesting duality properties \cite{gaiotto,gaiotto-maldacena, alday-gaiotto-tachikawa}. One can similarly compactify down to three \cite{dimofte-gaiotto-gukov} and to two \cite{gadde-gukov-putrov} dimensions. 

It is reasonable to expect similar phenomena with different six-dimensional CFT's. This might teach us something about the $(2,0)$ theory, but also about the dynamics of CFT's in lower dimensions. Perhaps the simplest generalization of the $(2,0)$ theory occurs when one introduces orbifold singularities \cite{intriligator-6d,blum-intriligator,intriligator-6d-II}; the study of their compactifications on Riemann surfaces is just starting \cite{gaiotto-razamat,franco-hayashi-uranga,hanany-maruyoshi}. From the holographic perspective, however, these theories are not very different from the $(2,0)$ theory: their dual is simply AdS$_7\times S^4/\zz_k$ \cite{ferrara-kehagias-partouche-zaffaroni,ahn-oh-tatar}. 

Nevertheless, an interesting further generalization can be obtained via NS5--D6--D8-brane systems \cite{hanany-zaffaroni-6d,brunner-karch}.\footnote{One can engineer six-dimensional field theories also in F-theory \cite{heckman-morrison-vafa,dhtv,heckman-morrison-rudelius-vafa}.} This class consists of $(1,0)$ SCFT's which are non Lagrangian, but which can be described by a quiver on a ``tensor branch''. Their holographic duals were found relatively recently: first numerically in \cite{afrt}, then analytically in \cite{10letter}. Their interpretation as the duals of the SCFT's described above was given in \cite{gaiotto-t-6d}. Up to orbifolds and orientifolds, these are the most general AdS$_7$ solutions in perturbative type II supergravity. 

Although the compactifications of these theories to lower dimensions are not yet known, they can already be studied holographically: the corresponding AdS$_5$ and AdS$_4$ solutions were found respectively in \cite{afpt} and \cite{rota-t}. These solutions are similar in spirit to the duals of the compactifications of the $(2,0)$ theory \cite{maldacena-nunez,pernici-sezgin,acharya-gauntlett-kim}: namely, AdS$_7$ gets replaced by AdS$_5\times \Sigma_2$ or AdS$_4\times \Sigma_3$, and the internal space gets distorted in a certain way. What is perhaps nicer than expected is that this distortion is ``universal''. Namely, even though there are infinitely many AdS$_7$ solutions, the map to obtain the AdS$_5$ and AdS$_4$ metric is always the same. Moreover, the two maps are very similar to each other: they differ only by the value of certain numerical factors.

In this paper, we greatly extend this universality. We promote the maps to a more general Ansatz, where AdS$_7$ gets replaced by any seven-dimensional metric $g_{\mu\nu}$, and the internal space gets distorted in a way that depends on a single scalar parameter $X$. This Ansatz in fact becomes nothing but a reduction to a seven-dimensional effective theory. Its bosonic fields are $X$ and $g_{\mu\nu}$ themselves, together with a three-form potential, and an SU(2) gauge field which is related to the fibration of the internal space over the seven external dimensions. 

This effective theory is the so-called minimal gauged supergravity in seven dimensions \cite{townsend-vannieuwenhuizen,mezincescu-townsend-vannieuwenhuizen}, which describes the dynamics of (a gauged version of) the gravity multiplet with sixteen supercharges. It is a subsector of the bigger ``maximal'' \cite{pernici-pilch-vannieuwenhuizen} theory, which describes the gravity multiplet with thirty-two supercharges and has gauge group SO(5). Both theories can be obtained \cite{lu-pope,nastase-vaman-vannieuwenhuizen2} as consistent truncations from eleven dimensions. 

Here we find that the minimal theory can also be obtained from massive IIA, in \emph{infinitely many ways}. In each of these reductions, the supersymmetric AdS$_7$ vacuum is one of the solutions in \cite{afrt,10letter}. This is perhaps surprising, but the idea is that, in reducing, we are only using the ordinary differential equation (ODE) that the internal geometry has to solve in the vacuum, and not the details of the individual solution. Moreover, since our reduction procedure consists in comparing equations of motion, we have a direct proof that these are all consistent truncations of massive IIA.

Thus we can uplift to massive IIA any solution of the seven-dimensional supergravity, in infinitely many ways. For example, the theory has AdS$_5\times \Sigma_2$ \cite{maldacena-nunez}\footnote{This solution was actually obtained in the maximal theory, with SO(5) gauge group, but it is possible to show that it survives in the minimal theory.} and AdS$_4\times \Sigma_3$ \cite{pernici-sezgin} solutions. They uplift to those of \cite{afpt,rota-t}. In this sense we are explaining and extending the universality noticed in those papers. Minimal gauged supergravity also has ``Renormalization Group (RG) flow'' solutions that connect the above backgrounds to the AdS$_7$ maximally supersymmetric vacuum. This shows conclusively that the solutions of \cite{afpt,rota-t} are indeed dual to compactifications on $\Sigma_2$ and $\Sigma_3$ of the six-dimensional $(1,0)$ SCFT's. 

Minimal gauged supergravity also admits AdS$_3 \times \Sigma_4$ solutions, preserving $\mathcal{N} = 1$ and $\mathcal{N} = 2$ supersymmetry. In the latter case $\Sigma_4$ is a K\"{a}hler--Einstein manifold of negative constant curvature, while in the former case $\Sigma_4$ is (a compact quotient of) hyperbolic space $\mathbb{H}^4$. The corresponding CFT duals are two-dimensional $(0,2)$ and $(0,1)$  SCFTs. 
Uplifting these solutions yields new AdS$_3$ solutions of massive IIA supergravity. On the field theory side, this implies that all the six-dimensional SCFT's of \cite{hanany-zaffaroni-6d,brunner-karch,gaiotto-t-6d} can be compactified on four-manifolds $\Sigma_4$ to produce two-dimensional SCFT's. 

Finally, minimal gauged supergravity has a second vacuum, which is not supersymmetric. This means that there are also non-supersymmetric analytical AdS$_7$ solutions in massive IIA. Although we will not discuss these solutions in this paper, it would be interesting to analyze them further, for example by comparing them with the numerical non-supersymmetric solutions of \cite{junghans-schmidt-zagermann}.

This paper is organized as follows. In section \ref{sec:7d}, we will review the seven-dimensional minimal gauged supergravity. In section \ref{sec:10d} we will review the IIA AdS$_7$ solutions found numerically in \cite{afrt} and analytically in \cite{afpt}, and their AdS$_5$ and AdS$_4$ compactifications. In section \ref{sec:red} we will perform the reduction from massive IIA to seven-dimensional minimal gauged supergravity. Finally, in section \ref{sec:sol} we will discuss some supersymmetric solutions to seven-dimensional minimal gauged supergravity, which thanks to our results can be lifted to supersymmetric massive IIA solutions. 


\section{Minimal gauged supergravity in seven dimensions} 
\label{sec:7d}

The bosonic fields of seven-dimensional minimal gauged supergravity \cite{townsend-vannieuwenhuizen} are the graviton, a triplet of one-forms $\A^i$, $i = 1,\,2,\,3$, transforming in the adjoint representation of SU(2), a scalar $\varphi$ and a three-form $\Athree$. The corresponding Lagrangian is\footnote{The scalar and the form fields of the original paper have being rescaled by a factor of $\frac{1}{\sqrt{2}}$ and the constant $h$ by a factor of $\frac{1}{4}$.} 
\begin{align}\label{7dLagrangian}
\mathcal{L} = 
R 
&- \tfrac{1}{2} *d\varphi \wedge d\varphi  
- V(\varphi) *1
- \tfrac{1}{2} e^{\frac{4}{\sqrt{10}}\varphi} * \Ffour \wedge \Ffour 
- \tfrac{1}{2} e^{-\frac{2}{\sqrt{10}}\varphi} *\Ftwo^i \wedge \Ftwo^i \\ \nonumber
&+ \tfrac{1}{2}\Ftwo^i \wedge \Ftwo^i \wedge \Athree - h \Ffour \wedge \Athree \ ,
\end{align}
where $V(\varphi)$ is the scalar potential
\begin{equation}
V(\varphi) = 
2 h^2 e^{-\frac{8}{\sqrt{10}}\varphi} 
- 4 \sqrt{2} h g e^{-\frac{3}{\sqrt{10}}\varphi} 
- 2g^2 e^{\frac{2}{\sqrt{10}}\varphi} \ . 
\end{equation}
$\Ftwo^i = d \A^i - \frac{1}{2} g \epsilon^{ijk} \A^j \wedge \A^k$ and $\Ffour = d \Athree$ are the field strengths of $\A^i$ and $\Athree$ respectively. $g$ is the gauge coupling constant whereas the constant $h$ is referred to as the topological mass. 

If $h/g > 0$ the scalar potential has two extrema: a maximum at $e^{-\frac{5}{\sqrt{10}}\varphi} = \frac{1}{2\sqrt{2}} \frac{g}{h}$ and a minimum at $e^{-\frac{5}{\sqrt{10}}\varphi} = \frac{1}{\sqrt{2}} \frac{g}{h}$; only the former is supersymmetric \cite{mezincescu-townsend-vannieuwenhuizen}.

There is a dual formulation of the theory with a two- instead of a three-form. In this case, the topological mass and the corresponding term in the Lagrangian are absent and the scalar potential has no critical points.
In \cite{chamseddine-sabra-7d} it was shown that this version can be embedded in ten-dimensional type I supergravity.

The fermionic fields are the gravitino $\psi_{\mu a}$, $\mu = 0, \dots, 6$ and the dilatino $\lambda_a$. They are symplectic-Majorana spinors transforming as SU(2) doublets; $a=1,\,2$ is the symplectic-Majorana/SU(2) index. The supersymmetry variations of the fermions read
\begin{subequations}\label{7dSusyVariations}
\begin{align}
\delta_\xi \psi_{\mu a} &= (\nabla_\mu + ig (\A_{\mu})_a{}^b) \xi_b 
+ \tfrac{i}{10\sqrt{2}} e^{-\frac{1}{\sqrt{10}}\varphi} 
\left( \gamma_\mu{}^{\alpha_1\alpha_2} - 8\delta_\mu{}^{\alpha_1} \gamma^{\alpha_2} \right)
({\Ftwo}_{\alpha_1\alpha_2})_a{}^b \xi_b \nonumber \\
&\hspace{1.5cm} + \tfrac{1}{160} e^{\frac{2}{\sqrt{10}}\varphi}
\left(\gamma_\mu{}^{\alpha_1\alpha_2\alpha_3\alpha_4} - \tfrac{8}{3} \delta_\mu{}^{\alpha_1} \gamma^{\alpha_2\alpha_3\alpha_3} \right)
{\Ffour}_{\alpha_1\alpha_2\alpha_3\alpha_4} \, \xi_a + m \gamma_\mu \xi_a \ , \\
\delta_\xi \lambda_a &= \tfrac{1}{2\sqrt{2}} \cancel{\partial}\varphi \xi_a 
- \tfrac{i}{\sqrt{10}} e^{-\frac{1}{\sqrt{10}}\varphi} (\cancel{\Ftwo})_a{}^b \xi_b
+ \tfrac{1}{2\sqrt{5}} e^{\frac{2}{\sqrt{10}}\varphi} \cancel{\Ffour} \xi_a 
- \sqrt{5}(m + \tfrac{h}{2} e^{-\frac{4}{\sqrt{10}}\varphi} ) \xi_a \ , 
\end{align}
\end{subequations}
where
\begin{equation}
m = - \frac{h}{10} e^{-\frac{4}{\sqrt{10}}\varphi} - \frac{g}{5\sqrt{2}} e^{\frac{1}{\sqrt{10}}\varphi} \ .
\end{equation}
Furthermore,
\begin{equation}
(\A)_a{}^b = \A^i (T^i)_a{}^b \ , \qquad
({\Ftwo})_a{}^b = \Ftwo^i (T^i)_a{}^b \ .
\end{equation}
$T^i = \frac{1}{2} \sigma^i$ are the generators of SU(2), $\sigma^i$ being the Pauli matrices. 

The slash of a $p$-form $\mathcal{F}_p$ is defined as
\begin{equation}
\cancel{\mathcal{F}_p} \equiv \frac{1}{p!} {\mathcal{F}_p}_{\alpha_1 \dots \alpha_p} \gamma^{\alpha_2 \dots \alpha_p} \ .
\end{equation}


\section{\texorpdfstring{AdS$_7$ solutions in massive IIA supergravity}{AdS(7) solutions in massive IIA supergravity} } 
\label{sec:10d}

In this section we review the IIA AdS$_7$ solutions of \cite{afrt}. These, according to our embedding, are the uplift of the supersymmetric AdS$_7$ vacuum of the seven-dimensional minimal gauged supergravity. We also discuss compactifications of these solutions to AdS$_4$ and AdS$_5$ \cite{10letter}; these will be instrumental in coming up with an appropriate reduction Ansatz in section \ref{sec:red}. 

\subsection{The solutions} 
\label{sub:sol}
While there are infinitely many AdS$_7$ solutions in IIA supergravity, they all share a few fundamental features. The internal space $M_3$ is an $S^2$-fibration over an interval, whose coordinate we call $r$. The $S^2$ shrinks at the two endpoints of this interval, so that $M_3$ has the topology of an $S^3$. Metric and fluxes can be written in terms of three functions: the dilaton, the warping, and one function $x$ related to the volume of the $S^2$. All three only depend on $r$:
\begin{equation}
	\phi= \phi(r) \ ,\qquad A= A(r) \ ,\qquad x= x(r) \ .
\end{equation}
The metric now reads 
\begin{equation}
	ds^2_{10} = e^{2A} ds^2_{\mathrm{AdS}_7} +  ds^2_{M_3} \ , \qquad ds^2_{M_3} = dr^2 + \tfrac{1}{16} e^{2A} (1-x^2) ds^2_{S^2} \ . 
\end{equation}
$ds^2_{\mathrm{AdS}_7}$ and $ds^2_{S^2}$ are unit radius metrics on AdS$_7$ and $S^2$. The expression of the Neveu--Schwarz flux is
\begin{equation}
H = -\left(6 e^{-A} + F_0 \, x e^{\phi}\right) \vol_{M_3} \ , \\
\end{equation}
where $F_0$ is the Romans mass and $\phi$ the dilaton. The expression for the Ramond--Ramond two-form flux is
\begin{equation}
F_2 = \frac{1}{16} e^{A-\phi} \sqrt{1-x^2}\left( F_0 \, e^{A+\phi} x - 4 \right)\vol_{S^2} \ . \\
\end{equation}

The functions $\phi(r)$, $A(r)$, $x(r)$ obey a system of ODEs:
\begin{subequations}\label{eq:oder}
\begin{align}
\frac{d\phi}{dr} &= \frac{1}{4} \frac{e^{-A}}{\sqrt{1-x^2}}\left(12x + (2x^2-5) F_0 e^{A+\phi}\right) \ , \\
\frac{dx}{dr} &= - \frac{1}{2} e^{-A} \sqrt{1-x^2}\left(4 + x F_0 e^{A+\phi}\right) \ , \\
\frac{dA}{dr} &= \frac{1}{4} \frac{e^{-A}}{\sqrt{1-x^2}}\left(4x - F_0 e^{A+\phi}\right)\ .
\end{align}
\end{subequations}
Originally, in \cite{afrt}, the AdS$_7$ solutions were found by integrating this system numerically. However, it was later found in \cite{afpt} that the solutions are determined by a single function $\beta(y)$ satisfying a single ODE:
\begin{equation}\label{eq:ode}
	(q^2)'= \frac29 F_0 \ ,\qquad q\equiv -\frac{4y\sqrt{\beta}}{\beta'}\ ,
\end{equation}
where the new variable $y$ is defined by $dr = \left(\frac{3}{4}\right)^2 \frac{e^{3A}}{\sqrt{\beta}} dy$, and a prime denotes differentiation with respect to $y$. Now $A$, $\phi$, and $x$ are determined by 
\begin{equation}\label{eq:Abeta}
\begin{split}
		e^{A}= \frac23 \left(-\frac {\beta'}y\right)^{1/4}&\ ,\qquad e^{\phi}=\frac{(-\beta'/y)^{5/4}}{12\sqrt{4 \beta - y \beta'}}\ ,\\
		x^2=&\frac{-y \beta'}{4 \beta-y \beta'}\ .
\end{split}
\end{equation}

The ODE (\ref{eq:ode}) can be readily solved analytically by writing it as $16 y^2\frac \beta{(\beta')^2} = \frac29 F_0 (y - \hat y_0)$, with $\hat y_0$ a constant; this can now be integrated by quadrature. Without D8-branes, the generic solution \cite[Sec.~5.6]{afpt} has two special points, corresponding to the presence of two stacks with $k_1$ and $k_2$ D6-branes (or one stack of D6-branes and an O6-plane). One special case happens where $F_0=0$: in this case $k_1=-k_2\equiv k$, and the solution is $\beta=\frac 4{k^2} (y-y_0^2)$. (This solution can also be obtained as a reduction from AdS$_7\times S^4$ in M-theory \cite[Sec.~5.1]{afrt}.) Another special case happens when $k_2=0$: here $\beta= \frac8{F_0}(y-y_0)(y+2y_0)^2$ \cite[Sec.~5.5]{afpt}. 

More solutions can be obtained by introducing D8-branes. In this case, the Romans mass flux $F_0$ jumps as one crosses the D8's, and correspondingly the metric is continuous but has a discontinuous first derivative, as one expects from a domain wall. The positions of the D8's are fixed by various flux quantization conditions. The metric can be obtained by gluing together pieces of the analytic solutions described earlier; this can be done in such a way as to avoid D6-branes, or as to include them, as one wishes. All in all, one has an infinite set of solutions; they are in one-to-one correspondence \cite{gaiotto-t-6d} with NS5--D6--D8 systems \cite{hanany-zaffaroni-6d,brunner-karch}. The corresponding SCFT$_6$'s are non-Lagrangian, but an effective description is known on their tensor branch. 

In any case, we will not need to know too many details about the classification of the most general solutions, since the reduction to seven dimensions will work much in the same way for all of them. This is roughly because we will only need to use (\ref{eq:oder}), and not the actual expressions for the solutions.


\subsection{Supersymmetry parameters} 
\label{sub:susy-vac}

All the solutions we just described are $\mathcal{N} = 1$ supersymmetric. The original method to find them used a formulation of the supersymmetry equations in terms of differential forms, where the spinors were never explicitly used. However, in order to compare supersymmetry in ten dimensions to supersymmetry in seven, in section \ref{subsec:susy_var} we will actually need the supersymmetry parameters, which were given in \cite{rota-t}:\footnote{They were also independently computed by I.~Bakhmatov (unpublished notes).}
\begin{equation}\label{10dSusyVac}
\epsilon_1 = (\xi \otimes \chi_1 + \xi^c \otimes \chi^c_1) \otimes \ket{\uparrow}
\ , \qquad
\epsilon_2 = (\xi \otimes \chi_2 - \xi^c \otimes \chi^c_2) \otimes \ket{\downarrow}
\ .
\end{equation}
Here $\xi$ is a Killing spinor on AdS$_7$, while $\ket{\uparrow}$ and $\ket{\downarrow}$ are eigenvectors of the Pauli matrix $\sigma^3$, with eigenvalues $+1$ and $-1$ respectively. The expressions for $\chi_1$ and $\chi_2$ are
\begin{equation}\label{vac-chi}
\chi_1 = -i e^{\frac{A}{2}} e^{-i\frac{\pi}{2}\sigma_3} e^{i\frac{\alpha}{2}\sigma^3}
\chi_{S^2} \ , \qquad
\chi_2 =  e^{\frac{A}{2}} e^{-i\frac{\alpha}{2}\sigma^3}
\chi_{S^2} \ ,
\end{equation}
where $\sin\alpha = x$ and $\chi_{S^2}$ is a Killing spinor on $S^2$. The superscript $^c$ denotes charge conjugation. The SU$(2)$ R-symmetry acts on the doublets
 $(\xi, \, {\xi}^c)^t$, $(\chi_1, \, \chi_1^c)^t$ and $(\chi_2, \, - \chi_2^c)^t$ in the fundamental representation.


\subsection{\texorpdfstring{Compactifications to AdS$_5$ and AdS$_4$}{Compactifications to AdS(5) and AdS(4)}} 
\label{sub:comp}

It is possible to compactify the AdS$_7$ solution on $\mathbb{H}^2$ or $\mathbb{H}^3$ to AdS$_5$ and AdS$_4$ respectively, by associating the functions that determine the solutions, via the map \cite{10letter}
\begin{equation}\label{map}
e^A \rightarrow X^{\frac{15}{4}} e^A \ , \qquad
r \rightarrow X^{\frac{5}{4}} r \ , \qquad
x \rightarrow \frac{x}{\sqrt{w}} \ ,
\end{equation}
where $X$ is a constant parameter, with the value $X = 1$ for the AdS$_7$ solution and $w \equiv X^5(1-x^2) + x^2$.\footnote{The dilaton transforms as $e^{\phi} \rightarrow X^{\frac{5}{4}} \frac{e^\phi}{\sqrt{w}}$.} The corresponding geometries read
\begin{align}\label{eq:dsX}
ds^2_{10} = X^{\frac{15}{2}} e^{2A} &ds^2_7 + X^{\frac{5}{2}} ds^2_{M_3} \ , \qquad ds^2_{M_3} = dr^2 + \frac{1-x^2}{16 w} e^{2A} Ds^2_{S^2} \ ,
\\[10pt] \nonumber
&ds^2_7= \left\{ \begin{aligned}
\displaystyle 
&{ds^2_{\mathrm{{\rm AdS}}_5} + \tfrac{1}{3}ds^2_{\mathbb{H}^2}}\\[8pt]
\displaystyle
&{ds^2_{\mathrm{AdS}_4} + \tfrac{4}{5}ds^2_{\mathbb{H}^3}}
\end{aligned}\right.
\ , \qquad \qquad 
X^5 = \left\{ \begin{aligned}
\displaystyle 
&{\frac{3}{4}}\\[7pt]
\displaystyle
&{\frac{5}{8}}
\end{aligned}\right. \ ,
\end{align}
where $ds^2_{\mathbb{H}^2}$ and $ds^2_{\mathbb{H}^3}$ are metrics of unit radius. The $S^2$ is fibered over $\mathbb{H}^2$ or $\mathbb{H}^3$, with the U$(1)$ spin connection of $\mathbb{H}^2$ twisting a U$(1)$ isometry inside the full SU$(2)$ isometry of $S^2$ in the first case and the SU$(2)$ spin connection of $\mathbb{H}^3$ twisting the whole isometry in the second. 

One can then quotient $\mathbb{H}^2$ and $\mathbb{H}^3$ by discrete subgroups of ${\rm PSL}(2,\R)$ and ${\rm PSL}(2,\C)$, so as to obtain respectively a Riemann surface $\Sigma_2$ of genus $g\ge 2$, or a compact hyperbolic manifold $\Sigma_3$. The holographic interpretation of these solutions is then similar to the familiar Maldacena--N\'u\~nez case \cite{maldacena-nunez}: they represent twisted compactifications of SCFT$_6$'s to SCFT$_4$'s and SCFT$_3$'s.  

The fact that both solutions can be written as (\ref{eq:dsX}) suggests a reduction Ansatz for massive IIA supergravity on $M_3$: promote $X$ to scalar field in seven dimensions and introduce seven-dimensional gauge vector fields gauging the SU(2) isometry of $M_3$. 



\section{Reduction} 
\label{sec:red}

In this section we present the Ansatz for the Kaluza-Klein reduction of massive IIA supergravity on $M_3$, to the seven-dimensional minimal gauged supergravity. Our approach to verifying the consistency of the reduction (or truncation) is to substitute the Ansatz into the ten-dimensional equations of motion and show that these are satisfied provided that the seven-dimensional equations of motion are satisfied. Vice versa, any solution of the lower-dimensional theory can be uplifted on $M_3$ to an exact solution of the higher-dimensional theory. This is described in subsection \ref{subsec:eom}.

In subsection \ref{subsec:susy_var} we take a further step and show that  any \emph{supersymmetric} solution of the seven-dimensional theory uplifts to a solution that also preserves supersymmetry. We provide a decomposition Ansatz for the ten-dimensional supersymmetry parameters and require that the supersymmetry variations of the fermion fields of IIA supergravity vanish. This condition yields a set of equations for the seven-dimensional part of the supersymmetry parameters: it is exactly the set of equations one obtains by setting to zero the supersymmetry variations of the fermion fields of the seven-dimensional minimal gauged supergravity. Vice versa, any spinor $\xi_a$ such that the lower-dimensional supersymmetry transformations \eqref{7dSusyVariations} vanish can be uplifted so that the higher-dimensional supersymmetry transformations vanish as well.

\subsection{Equations of motion}
\label{subsec:eom}

The Ansatz for the ten-dimensional metric is 
\begin{equation}\label{10dMetric}
\ell^{-1} ds^2_{10} = \tfrac{1}{8} g^2 X^{-\frac{1}{2}} e^{2A}ds^2_7 + X^{\frac{5}{2}} ds^2_{M_3} \ , \qquad ds^2_{M_3} = dr^2 + \frac{1-x^2}{16w}e^{2A} Ds^2_{S^2} \ , 
\end{equation}
where $\ell \equiv \frac{8\sqrt{2}}{g^3}$ and
\begin{equation}
w \equiv X^5(1-x^2) + x^2  \ .
\end{equation}
The parameter $X$ is promoted in this section to a scalar in
seven dimensions; it will turn out to be related to the scalar
$\varphi$ of section \ref{sec:7d}. It was a constant for the AdS solutions of (\ref{eq:dsX}). The covariantized metric $Ds^2_{S^2}$ on the two-sphere is 
\begin{equation}
Ds^2_{S^2} \equiv Dy^iDy^i \ , \qquad Dy^i \equiv dy^i + \epsilon^{ijk} y^j g \A^k \ .
\end{equation}
$y^i$ parametrize $S^2 \in \mathbb{R}^3$ as the locus $y^iy^i = 1$; explicitly
\begin{equation}
y^i = (\sin\theta\cos\psi, \, \sin\theta\sin\psi, \, \cos\theta) \ .
\end{equation}
In angular coordinates, $Ds^2_{S^2}$ reads
\begin{equation}
Ds^2_{S^2} = 
(d\theta + K^\theta_i g \A^i )^2 + 
\sin^2\theta(d\psi + K^\psi_i g \A^i)^2 \ ,
\end{equation} 
where $K_1 = \cot\theta \cos\psi \partial_\psi + \sin\psi \partial_\theta$, $K_2 = \cot\theta \sin\psi \partial_\psi - \cos\psi \partial_\theta$ and $K_3 = - \partial_\psi$ are the Killing vectors generating the SO(3) isometry of $S^2$. 

The Ansatz for the dilaton $\Phi$ is 
\begin{equation}
e^{2\Phi} = \ell \frac{X^{\frac{5}{2}}}{w} e^{2\phi} \ .
\end{equation}
Here and in what follows, $\phi$ is the dilaton for the AdS$_7$ solution presented in section \ref{sub:sol}.

The Ansatz for the Neveu-Schwarz potential $B$ is 
\begin{equation}
\ell^{-1} B = 
\frac{1}{16} e^{2A} \frac{x \sqrt{1-x^2}}{w} \vol_2 
- \frac{1}{2} e^A dr \wedge  (a - \tfrac{1}{2}y^i\A^i) \ , 
\end{equation}
where $\vol_2 \equiv \frac{1}{2} \epsilon^{ijk} y^i Dy^{jk}$ is the volume of the covariantized $S^2$ and $a$ is defined via $da = - \frac{1}{2} \vol_{S^2}$. $H=dB$ then reads
\begin{align}
\ell^{-1} H &= 
\left\{(2-6X^5+4X^{10})x^2 -2X^5 -4 X^{10} \right\} w^{-1} e^{-A} \vol_{M_3} - X^5 w^{-1} \ell F_0 \, e^{\phi} x\vol_{M_3} \nonumber \\
&-\frac{1}{4} e^A dr \wedge y^i g \mathcal{F}_2^i - \frac{1}{16} w^{-1}  e^{2A} x \sqrt{1-x^2} g \mathcal{F}_2^i \wedge Dy^i
-\frac{5}{16} X^4 w^{-2} e^{2A} x (1-x^2)^{\frac{3}{2}} dX \wedge \vol_2 \ .
\end{align}

The Ans\"{a}tze for the Ramond-Ramond fluxes are
\begin{subequations}
\begin{align}
F_2 &= - q \left( \vol_2 + y^ig\mathcal{F}_2^i\right) + \frac{1}{16} w^{-1} \ell F_0 \, e^{2A} x \sqrt{1-x^2} \vol_2 \ ,\\ 
\ell^{-1} F_4 &= - \frac{q}{16} w^{-1} e^{2A} x \sqrt{1-x^2} y^i g\mathcal{F}_2^i \wedge \vol_2 -\frac{q}{4} e^{A} dr \wedge \epsilon^{ijk} g \mathcal{F}_2^i \wedge y^j Dy^k \\ \nonumber &- \frac{q}{2} e^{A} dr \wedge X^4 g^2 *_7 \mathcal{F}_4
- \ell^{-1} \frac{1}{2} e^{3A-\phi} x \mathcal{F}_4 \ , 
\end{align}
\end{subequations}
where $q \equiv \frac{1}{4} e^{A-\phi}\sqrt{1-x^2}$. $F_2$ and $F_4$ must obey the Bianchi identities
\begin{equation}
dF_2 - H F_0 = 0 \ , \qquad
dF_4 - H \wedge F_2 = 0\ .
\end{equation}
A way to see that this is the case for the above expressions is to note that
\begin{subequations}
\begin{align}
F_2 - B F_0 &= dC_1 \ , \\
F_4 - \frac{1}{2F_0} F_2 \wedge F_2 &= dC_3  \label{dC3} \ ,
\end{align}
\end{subequations}
where
\begin{subequations}
\begin{align}
C_1 &= 2q(a - \tfrac{1}{2} y^i\A^i)  \ , \\
C_3 &= - \frac{q^2}{2F_0} (\epsilon^{ijk} g \mathcal{F}_2^i y^j Dy^k + g^2 \omega_3)
- \frac{1}{2} e^{3A-\phi} x \mathcal{A}_3 \ .
\end{align}
\end{subequations}
$\omega_3 \equiv \A^i \wedge \Ftwo^i + \frac{1}{6} g \epsilon^{ijk} \A^i \wedge \A^j \wedge \A^k$, satisfying $d\omega_3 = \Ftwo^i \wedge \Ftwo^i$. In deriving $\eqref{dC3}$ one has to take into account the ``odd-dimensional self-duality" equation \cite{pilch-townsend-vannieuwenhuizen}
\begin{equation}\label{oddselfduality}
X^4 *_7 \mathcal{F}_4 = - \tfrac{1}{\sqrt{2}} g \mathcal{A}_3 + \tfrac{1}{2} \omega_3 \ .
\end{equation} 

The next step is to obtain the equations that the seven-dimensional fields satisfy, by substituting the Ans\"{a}tze for the ten-dimensional fields into the equations of motion of IIA supergravity. 

We employ the democratic formulation \cite{democratic} of type II supergravity and work in the string frame. The equations of motion of the fluxes are
\begin{equation}
(d + H \wedge) * F = 0 \ , \qquad
d(e^{-2\Phi} * H) - \tfrac{1}{2} \sum_p * F_{p} \wedge F_{p-2} = 0 \ ,
\end{equation}
where $F \equiv \sum_{p=0,2,4,6,8,10} F_p$. The Einstein equations are
\begin{equation}
R_{MN} + 2 \nabla_M \nabla_N \Phi - \tfrac{1}{2} H_M \cdot H_N - \tfrac{1}{4} e^{2\Phi} F_M \cdot F_N = 0 \ .
\end{equation}
where $F_M \cdot F_N \equiv \frac{1}{(p-1)!}\sum_p {F_p}_M{}^{M_1 \dots M_{(p-1)}} {F_p}_{N M_1 \dots M_{(p-1)}}$ and similarly for $H_M \cdot H_N $. Finally the dilaton equation is
\begin{equation}
\nabla^2 \Phi - (\nabla\Phi)^2 + \tfrac{1}{4} R - \tfrac{1}{8} H^2 = 0 \ .
\end{equation}

Substituting the Ans\"{a}tze into the flux and dilaton equations of motion, we arrive at the following equations for the seven-dimensional fields:
\begin{subequations}\label{7dEquationsOfMotion}
\begin{align}
\label{7dX}
0 &= d(X^{-1}*_7dX) + \tfrac{1}{5} g^2(X^{-8}-3X^{-3}+2X^2)\vol_7 \\ \nonumber
& \hspace{3cm} - \tfrac{1}{5}X^{4} *_7  \mathcal{F}_4 \wedge \mathcal{F}_4 + \tfrac{1}{10} X^{-2} *_7 \Ftwo^i \wedge \Ftwo^i   \ , \\
\label{7dF4}
0 &= d(X^4 *_7\mathcal{F}_4) + \tfrac{1}{\sqrt{2}} g \mathcal{F}_4 - \tfrac{1}{2} \Ftwo^i \wedge \Ftwo^i  \ , \\
\label{7dF2}
0 &= D(X^{-2} *_7 \Ftwo^i) - \Ftwo^i \wedge \mathcal{F}_4  \ .
\end{align}
\end{subequations}
In particular, \eqref{7dF4} and \eqref{7dF2} come from the equations of motion of $F_4$ and $F_2$ respectively, while both equations of motion of $H$ and $\Phi$ give rise to \eqref{7dX}. 

In order to reduce the Einstein equations, we compute the Riemann and subsequently the Ricci tensor via the curvature two-form $R^A{}_B = d \omega^A{}_B + \omega^A{}_C \wedge \omega^C{}_B$; the spin connection $\omega^A{}_B$ is that of the orthonormal frame introduced in appendix \ref{sec:frame}.
After a lengthy calculation we find that the ten-dimensional Einstein equations, upon using \eqref{7dX}, reduce to 
\begin{equation}\label{7dEinsteinEquations}
\begin{split}
R_{\mu\nu} &- 5 X^{-2} \partial_\mu X \partial_\nu X 
  - \tfrac{1}{20} g^2 \left(X^{-8} - 8 X^{-3} - 8 X^2\right) g_{\mu\nu} \\ 
& - \tfrac{1}{2} X^{-2} \left({\Ftwo^i}_\mu \cdot {\Ftwo^i}_\nu - \tfrac{1}{5} {\Ftwo^i}^2 g_{\mu\nu}\right) 
  - \tfrac{1}{2}X^4\left({\mathcal{F}_4}_\mu \cdot {\mathcal{F}_4}_\nu - \tfrac{3}{5} \mathcal{F}_4^2 g_{\mu\nu}\right) = 0 \ .
\end{split}
\end{equation} 

Equations \eqref{7dEquationsOfMotion} and \eqref{7dEinsteinEquations} can be derived from the Lagrangian \eqref{7dLagrangian} for 
\begin{equation}
	X = e^{\frac{1}{\sqrt{10}}\varphi} \ ,\qquad h = \frac{g}{2\sqrt{2}} \ .
\end{equation}

\subsection{Supersymmetry}
\label{subsec:susy_var}

The supersymmetry transformations of the gravitini of IIA supergravity are
\begin{equation}\label{10dSusyVariationsG}
\delta {\Psi_1}_M = \left(\nabla_M - \tfrac{1}{4} H_M \right) \epsilon_1 - \tfrac{1}{16} e^{\Phi}  F \Gamma_M
\epsilon_2 \ , \qquad
\delta {\Psi_2}_M = \left(\nabla_M + \tfrac{1}{4} H_M \right) \epsilon_2 - \tfrac{1}{16} e^{\Phi} \lambda(F) \Gamma_M
 \epsilon_1 \ .
\end{equation}
Fermion fields with a subscript $1$  have positive chirality, whereas fermion fields with a subscript $2$ have negative chirality. The suppressed indices of the fluxes are contracted with anti-symmetric products of gamma matrices. $\lambda$ is an operator acting on a $p$-form as $\lambda(F_p) = (-1)^{\left[\frac{p}{2}\right]} F_p$, where the square brackets denote the integer part of $\frac{p}{2}$. 
The supersymmetry transformations of the dilatini are 
\begin{equation}\label{10dSusyVariationsD}
\delta \lambda_1 = \left(\partial \Phi - \tfrac{1}{2} H \right) \epsilon_1 - \tfrac{1}{16} e^{\Phi} \Gamma^M F \Gamma_M
\epsilon_2 \ , \qquad 
\delta \lambda_2 = \left(\partial \Phi + \tfrac{1}{2} H \right) \epsilon_2  - \tfrac{1}{16} e^{\Phi} \Gamma^M \lambda(F) \Gamma_M
 \epsilon_1 \ .   
\end{equation}

The decomposition Ansatz for the ten-dimensional supersymmetry parameters is 
\begin{equation}\label{10dSusyParameters}
\epsilon_1 = (\xi \otimes \chi_1 + \xi^c \otimes \chi^c_1) \otimes \ket{\uparrow}
\ , \qquad
\epsilon_2 = (\xi \otimes \chi_2 - \xi^c \otimes \chi^c_2) \otimes \ket{\downarrow}
\ .
\end{equation}
This is analogous to \eqref{10dSusyVac}, but now $\xi$ is a generic seven-dimensional spinor, rather than a Killing one; the symplectic-Majorana doublet $\xi_a$ is  $(\xi, \, {\xi}^c)^t$.   The expressions for $\chi_1$ and $\chi_2$ are formally identical to \eqref{vac-chi},
\begin{equation}
\chi_1 = -i e^{\frac{A}{2}} e^{-i\frac{\pi}{2}\sigma_3} e^{i\frac{\alpha}{2}\sigma^3}
\chi_{S^2} \ , \qquad
\chi_2 =  e^{\frac{A}{2}} e^{-i\frac{\alpha}{2}\sigma^3}
\chi_{S^2} \ ;
\end{equation}
however, $\sin\alpha$ deviates from its vacuum value, $\sin\alpha = x$, following the map \eqref{map}: i.e. $\sin\alpha = w^{-\frac{1}{2}} x$. Accordingly, $\cos\alpha \equiv w^{-\frac{1}{2}} X^{\frac{5}{2}} \sqrt{1-x^2}$.

We can decompose the ten-dimensional supersymmetry transformations by splitting Cliff$(1,9)$ as\footnote{$\alpha = 0,\dots,6$, $a = 1,\,2,\,3$.}
\begin{equation}
\Gamma_\alpha = \gamma_\alpha \otimes \mathbb{I} \otimes \sigma_2
\ , \qquad
\Gamma_{a+6} = \mathbb{I} \otimes \sigma_a \otimes \sigma_1 \ ,
\end{equation}
and substituting for \eqref{10dSusyParameters}. Setting \eqref{10dSusyVariationsD} to zero amounts to
\begin{equation}
0 = \tfrac{5}{2} X^{-1} \cancel{\partial} X \xi_a + \tfrac{1}{2} X^2 \cancel{\mathcal{F}_4} \xi_a - \tfrac{i}{\sqrt{2}} X^{-1} (\cancel{\Ftwo}^i)_a{}^b \xi_b - \tfrac{1}{\sqrt{2}} g(X^{-4} - X) \xi_a  \ ,
\end{equation} 
whereas setting \eqref{10dSusyVariationsG} to zero amounts to the above equation for the internal components and
\begin{align}
0 &= (\nabla_\mu + i g ({\A^i}_\mu)_a{}^b) \xi_b 
+ \tfrac{i}{10\sqrt{2}} X^{-1} 
\left( \gamma_\mu{}^{\alpha_1\alpha_2} - 8\delta_\mu{}^{\alpha_1} \gamma^{\alpha_2} \right)
({\Ftwo^i}_{\alpha_1\alpha_2})^a{}_b \xi_b \nonumber \\ 
&+ \tfrac{1}{160} X^2
\left(\gamma_\mu{}^{\alpha_1\alpha_2\alpha_3\alpha_4} - \tfrac{8}{3} \delta_\mu{}^{\alpha_1} \gamma^{\alpha_2\alpha_3\alpha_4} \right)
{\mathcal{F}_4}_{\alpha_1\alpha_2\alpha_3\alpha_4} \, \xi_a 
- g\left(\tfrac{1}{20\sqrt{2}} X^{-4} + \tfrac{1}{5\sqrt{2}} X\right)\gamma_\mu \xi_a  \ .
\end{align}
for the external ones.
These constraints on $\xi_a$ are no other than those that one obtains by setting \eqref{7dSusyVariations} to zero, for $X = e^{\frac{1}{\sqrt{10}}\varphi}$ and $h = \frac{g}{2\sqrt{2}}$.

Thus, preserved supersymmetry in seven dimensions guarantees preserved supersymmetry in ten.


\section{Solutions: compactifications and flows} 
\label{sec:sol}

In this section we discuss (supersymmetric) anti-deSitter solutions of seven-dimensional minimal gauged supergravity,\footnote{The parameter $h$ is set equal to $\frac{g}{2\sqrt{2}}$, in accordance to the result of the reduction presented in the previous section.} along with holographic renormalization group (RG) flows, interpolating between the supersymmetric AdS$_7$ vacuum and lower-dimensional anti-deSitter vacua. All these uplift to massive IIA in ten dimensions via the formulas presented in the previous section. In particular, 
we consider the AdS$_5$ and AdS$_4$ solutions which uplift  to the ten-dimensional ones reviewed in section \ref{sub:comp}, and more notably, AdS$_3$ solutions which uplift to \emph{new} AdS$_3$ solutions of massive IIA supergravity with $\mathcal{N} = 1$ and $\mathcal{N} = 2$ supersymmetry.

\subsection{\texorpdfstring{AdS$_5$ and AdS$_4$}{AdS(5) and AdS(4)}} 
\label{sub:45}

$\mathcal{N} = 1$ and $\mathcal{N} = 2$ supersymmetric AdS$_5 \times \mathbb{H}^2$ solutions were first found in \cite{maldacena-nunez}, in a certain truncation of the maximal gauged supergravity in seven dimensions, keeping two scalars and two U$(1)$ gauge vector fields. In the case of the ${\cal N}=1$ solution, the two scalars and the two gauge vector fields are set to be equal and thus, the solution can also be embedded in the minimal theory of section \ref{sec:7d}.\footnote{The translation between the languages of \cite[appendix 7.3]{maldacena-nunez} and section \ref{sec:7d} is: $m \equiv \frac{g}{\sqrt{2}}$, $\lambda_1 = \lambda_2 = -\phi/2 \equiv \frac{\varphi}{2\sqrt{10}}$.}

The AdS$_5 \times \mathbb{H}^2$ geometry is a subset of  warped product geometries
\begin{equation}\label{eq:RG5}
ds^2_7 = e^{2f_1(r)} (dr^2 + ds^2_{\mathbb{R}^{3,1}}) + e^{2f_2(r)} ds^2_{\mathbb{H}^2} \ ,
\end{equation}
with a boundary condition for $f_1$ and $f_2$ as $r \to 0$, $f_1 \sim f_2 \sim \log r$. That is, asymptotically or in the UV the metric approaches AdS$_7$ with an $\mathbb{R}^{3,1} \times \mathbb{H}^2$ boundary. In order to preserve supersymmetry, the U$(1)$ gauge field is identified with the spin connection of $\mathbb{H}^2$  while $f_1$ and $f_2$ (as well as the scalar) are subject to a set of ODEs --- these can be found in \cite[Eq.~(27)]{maldacena-nunez}. 

The latter admit an AdS$_5 \times \mathbb{H}^2$ solution, which (in our language) reads
\begin{equation}\label{eq:34}
ds^2_7 = \frac{8}{g^2} e^{\frac{8}{\sqrt{10}}\varphi} 
\left( ds^2_{\AdS_5} + \tfrac{1}{3} ds^2_{\mathbb{H}^2} \right) \ , \qquad e^{\frac{5}{\sqrt{10}}\varphi} = \frac{3}{4} \ ,
\end{equation}
with the field strength of the U$(1)$ gauge field $g\Ftwo^i = - \vol_{\mathbb{H}^2} \, \delta^{i3}$, while the three-form potential is equal to zero. In \cite{bah-beem-bobev-wecht}, it was shown numerically (within a broader context) that the AdS$_5 \times \mathbb{H}^2$ solution arises as the IR fixed point of an RG flow that connects it to the AdS$_7$ region.

An $\mathcal{N} = 1$ supersymmetric AdS$_4 \times \mathbb{H}^3$ solution of seven-dimensional minimal gauged supergravity was first found in \cite{pernici-sezgin}. 
The metric and the scalar field of the solution read
\begin{equation}\label{eq:58}
ds^2_7 = \frac{8}{g^2} e^{\frac{8}{\sqrt{10}}\varphi} 
\left( ds^2_{\AdS_4} + \tfrac{4}{5} ds^2_{\mathbb{H}^3} \right) \ , \qquad e^{\frac{5}{\sqrt{10}}\varphi} = \frac{5}{8} \ .
\end{equation}
The SU$(2)$ gauge field is identified with the SU(2) spin connection $\omega^{ij}$ of $\mathbb{H}^3$ via
\begin{equation}
g \A^i = \tfrac{1}{2} \epsilon^{ijk} \omega^{jk} \ .
\end{equation}
The field strength is then $g\Ftwo^i = \tfrac{1}{2} \epsilon^{ijk} R^{jk}$, where $R^{jk}$ is the curvature two-form of the spin connection, while the three-form potential is zero.

It was later shown numerically \cite{acharya-gauntlett-kim} --- in an analogous analysis to that for the AdS$_5 \times  \mathbb{H}^2$ solution ---  that this solution also arises as the IR fixed point of an ``RG flow geometry", 
\begin{equation}\label{eq:RG4}
ds^2_7 = e^{2f_1(r)} (dr^2 + ds^2_{\mathbb{R}^{2,1}}) + e^{2f_2(r)} ds^2_{\mathbb{H}^3} \ ,
\end{equation}
with $f_1 \sim f_2 \sim \log r$ in the UV and the corresponding values for the AdS$_4 \times \mathbb{H}^3$ solution in the IR. 

The existence of the above RG flow solutions in the seven-dimensional minimal gauged supergravity, in conjunction with the consistent truncation of massive IIA supergravity presented in this paper, shows that the AdS$_5$ and AdS$_4$ solutions of \cite{afpt,rota-t} are connected to the AdS$_7$ ones of \cite{afrt} by RG flows. This proves that the solutions of \cite{afpt,rota-t} are dual to compactifications of six-dimensional $(1,0)$ theories on $\Sigma_2$ and $\Sigma_3$ manifolds of negative curvature.


\subsection{\texorpdfstring{AdS$_3$}{AdS(3)}} 
\label{sub:ads3}

We now turn to the supersymmetric AdS$_3$ solutions. The first one is AdS$_3 \times \mathbb{H}^4$ preserving two (real) supercharges. The metric and the scalar field of the solution read
\begin{equation}\label{eq:712}
ds^2_7 = \frac{2}{g^2} e^{-\frac{2}{\sqrt{10}}\varphi} 
\left( ds^2_{\AdS_3} + \tfrac{4}{7} ds^2_{\mathbb{H}^4} \right) \ , \qquad e^{\frac{5}{\sqrt{10}}\varphi} = \frac{7}{12} \ .
\end{equation}
The SU$(2)$ gauge field equals the self-dual part of the SO(4) spin connection of $\mathbb{H}^4$.
\begin{equation}\label{eq:A712}
g \A^i = \tfrac{1}{2} \epsilon^{ijk} \omega^{jk} + \omega^{i4} \ .
\end{equation}
The field strength is then $g\Ftwo^i = \tfrac{1}{2} \epsilon^{ijk} R^{jk} + R^{i4}$. Finally, the four-form flux is proportional to the volume of $\mathbb{H}^4$:
\begin{equation}
\Ffour =  \frac{3 \sqrt{2}}{g^3} \vol_{\mathbb{H}^4} \ .
\end{equation}

The second one is AdS$_3 \times M_4$, where $M_4$ is K\"{a}hler--Einstein of constant negative curvature $-4$ (for example $\mathbb{H}^2 \times\mathbb{H}^2$), preserving four supercharges. The metric and the scalar field of the solution read
\begin{equation}\label{eq:43}
ds^2_7 = \frac{2}{g^2} e^{-\frac{2}{\sqrt{10}}\varphi} 
\left( ds^2_{\AdS_3} + \tfrac{4}{3} ds^2_{M_4} \right) \ , \qquad e^{\frac{5}{\sqrt{10}}\varphi} = \frac{4}{3}\ .
\end{equation}

Only a U$(1) \subset$ SU$(2)$ gauge field is non-zero and is identified with the center U$(1)$ component of the  U$(2)$ spin connection of $M_4$, or equivalently with the K\"{a}hler connection on the canonical bundle of $M_4$. Taking the spin connection of the center U$(1)$ to be the truncation of the self-dual part of the spin connection we can write
\begin{equation}
g \A^i = (\omega^{12} + \omega^{34}) \delta^{i3}\ .
\end{equation}
The field strength is then identified with the Ricci form of $M_4$. Finally, the four-form flux is proportional to the volume of $M_4$:
\begin{equation}
\Ffour =  \frac{\sqrt{2}}{g^3} \vol_{M_4} \ .
\end{equation}

The above AdS$_3$ solutions were also found in \cite{gauntlett-kim-waldram} as the IR fixed points of RG flows constructed in certain truncations of the maximal seven-dimensional gauged supergravity. When uplifted to M-theory, the AdS$_3 \times M_4$ solution arises from M5-branes wrapping K\"{a}hler four-cycles in Calabi--Yau four-folds while the AdS$_3 \times \mathbb{H}^4$ one from M5-branes wrapping Cayley four-cycles in manifolds of Spin(7) holonomy. The scalar and gauge field sector of the truncations can be identified with the corresponding ones of the minimal theory, while the three-form potential sector is formulated in a dual frame, via \eqref{oddselfduality}. The AdS$_3 \times M_4$ solution was also constructed with different methods in \cite{cariglia-macconamhna}.

Let us conclude with a few words on the field theory duals of the solutions we described in this section. In the first case, (\ref{eq:712}), the SU(2) R-symmetry of the original AdS$_7$ solution is completely broken by the gauge fields (\ref{eq:A712}). Since no R-symmetry is left, the dual field theory should be a two-dimensional $(0,1)$ SCFT. In the second case, (\ref{eq:43}), only a U(1) gauge field is switched on; its commutant in SU(2)$_{\rm R}$ is the U(1) itself. This signals that the IIA uplift still has a U(1) isometry; this is the R-symmetry of the dual theory, which should then be a $(0,2)$ SCFT$_2$ this time. It would be interesting to study these theories, perhaps generalizing \cite{gadde-gukov-putrov}. 

We can also use AdS/CFT to compute the number of degrees of freedom in these theories, along the lines of \cite[Sec.~5.8]{afpt}, \cite[Sec.~4.8]{rota-t}. In fact, the formalism in this paper allows us to write a general formula. Let ${\cal F}_{0,2}$ be the coefficient in the scaling of the free energy ${\cal F}_2 = {\cal F}_{0,2}T^2 V$ with temperature $T$ and volume $V$, for a SCFT in $2$ dimensions. Then, the coefficient ${\cal F}_{0,6}$ for an $(1,0)$ theory dual to massive IIA and the coefficient for a theory obtained by compactifying it on a $4$-dimensional space $\Sigma_4$ are related by
\begin{equation}
	\frac{{\cal F}_{0,2}}{{\cal F}_{0,6}}= \frac{1}{(2 X_{\rm IR})^5} {\rm Vol}(\Sigma_4) \ ,
\end{equation}
where $X_{\rm IR}$ is the value of $X$ for the lower-dimensional AdS solution (recall that $X=e^{\frac1{\sqrt{10}}\varphi}$).\footnote{The corresponding formula for both the AdS$_4$ and AdS$_5$ solutions is 
\begin{equation}
	\frac{{\cal F}_{0,6-d}}{{\cal F}_{0,6}}= \left(X_{\rm IR}\right)^{20}  {\rm Vol}(\Sigma_d) \ , \qquad d = 3,2 \ .
\end{equation}
}
For example, for the solution (\ref{eq:43}), we get ${\cal F}_{0,2}/{\cal F}_{0,6} = 1/2^5 \cdot 3/4 \,  {\rm Vol}(\Sigma_4)$.


\section{Concluding remarks} 
\label{sec:concl}

We have constructed a consistent truncation of massive IIA supergravity on $M_3$, to seven-dimensional minimal gauged supergravity, where $M_3$ is the internal manifold of the AdS$_7$ solutions of \cite{afrt, 10letter}. The truncation is universal: it applies to the whole infinite family of Riemannian metrics on $M_3$. These exhaust the supersymmetric AdS$_7$ backgrounds of IIA supergravity. The outcome of this truncation is that any solution of the seven-dimensional theory uplifts to a solution of massive IIA supergravity in ten dimensions. Working at the level of the supersymmetry variations, we have also showed that supersymmetry is preserved in this process.

As an application of our result, we focused on RG flows in seven dimensions, which in ten dimensions connect the AdS$_5$ and AdS$_4$ solutions of \cite{afpt} and \cite{rota-t} to the AdS$_7$ ones. Furthermore, AdS$_3$ vacua in seven dimensions produce new $\mathcal{N} = 1$ and $\mathcal{N} = 2$ supersymmetric AdS$_3$ solutions of massive IIA supergravity, dual to $(0,1)$ and $(0,2)$ SCFT's in two dimensions. This is an addition to the series of compactifications of the AdS$_7$ backgrounds to five and four dimensions. 

In \cite{gaiotto-t-6d} it was argued that the AdS$_7$ solutions of massive IIA supergravity are the gravity duals of six-dimensional $(1,0)$ SCFT's, engineered by NS5--D6--D8-brane intersections \cite{hanany-zaffaroni-6d, brunner-karch}. The universal character of the present truncation implies that supergravity in seven dimensions describes a sector common to all these theories --- including also the $(2,0)$ theory itself, described by the original M-theory reduction of \cite{lu-pope}.

A similar ``common sector'' phenomenon is witnessed in five dimensions, where it was found that for every AdS$_5$ solutions of IIB there is a consistent truncation down to minimal five-dimensional supergravity \cite{gauntlett-varela}.\footnote{For an earlier example, concerning a reduction from M-theory, see \cite{Gauntlett:2006ai}.} In the same paper, it was conjectured that this phenomenon should hold in any dimensions; our results prove their conjecture in dimension seven. For certain internal manifolds, it is possible to excite more modes and get bigger theories, e.g.~for Sasaki--Einstein reductions \cite{gauntlett-kim-varela-waldram}.

Beyond this common sector, discerning finer differences between the CFT$_6$'s would require more sophisticated reduction procedures, where one keeps more internal modes. These might be gravity modes, or they could come from the D6- and D8-branes which are present in all the IIA vacua of \cite{afrt, 10letter}. In both cases, one would end up coupling the minimal theory to vector multiplets.\footnote{\cite{danielsson-dibitetto-fazzi-vanriet} argues however that the massive IIA vacua cannot be truncated either to the maximal theory, with gauge group SO(5), nor to a theory with gauge group SO(4) \cite{salam-sezgin} (which can be obtained as reduction from M-theory \cite{cvetic-lu-pope-sadrzadeh-tran,karndumri}).} 

Via the gauge/gravity duality, our work paves the way for a broader study of the aforementioned six-dimensional field theories. Asymptotically locally anti-deSitter solutions of seven-dimensional gauged supergravity can probe regions away from the superconformal fixed point. The Kaluza--Klein spectrum of the AdS$_7 \times M_3$ backgrounds, beyond the massless modes, can be used to analyze the spectrum of the dual operators. Finally, since the minimal seven-dimensional gauged supergravity can also be embedded in M-theory\cite{lu-pope}, lessons learned from the more extensively studied AdS$_7$/CFT$_6$ correspondence stemming from the dynamics of M5-branes can guide us in the study of its $(1,0)$ cousin in the massive IIA theory.


\section*{Acknowledgments}
We would like to thank G.~Dibitetto and O.~Varela for interesting discussions. We are supported in part by INFN. A.P.~and A.T.~are also supported by the European Research Council under the European Union's Seventh Framework Program (FP/2007-2013) -- ERC Grant Agreement n. 307286 (XD-STRING). The research of A.T.~is also supported by the MIUR-FIRB grant RBFR10QS5J ``String Theory and Fundamental Interactions''.

\appendix

\section{Orthonormal frame and spin connection} 
\label{sec:frame}

We introduce the following orthonormal frame for the ten-dimensional metric \eqref{10dMetric}:
\begin{align}
e^\alpha =  \ell^{\frac{1}{2}} X^{-\frac{1}{4}} e^{A} \tilde{e}^\alpha \ , &\qquad
e^3 = \ell^{\frac{1}{2}} X^{\frac{5}{4}} dr \ , \\ \nonumber 
e^2 = \ell^{\frac{1}{2}} X^{\frac{5}{4}} e^f \sin\theta(d\psi + K^\psi_i g \A^i) \ , &\qquad
e^1 = \ell^{\frac{1}{2}} X^{\frac{5}{4}} e^f (d\theta + K^\theta_i g \A^i ) \ ,       
\end{align}
where $\alpha = 0, \dots, 6$ and $\tilde{e}^\alpha$ is the orthonormal frame for $ds^2_7$. Furthermore,
\begin{equation}
e^f \equiv  e^{A} \frac{1}{4} \sqrt{\frac{1-x^2}{w}} \ . 
\end{equation}
The spin connection of the frame is
\begin{subequations}
\begin{align}
&\omega^\alpha{}_\beta = 
{\tilde{\omega}}^\alpha{}_\beta 
- \frac{1}{2} e^{[\alpha} X^{-1} \partial_{\beta]} X
- \frac{1}{2} \ell^{\frac{1}{2}} X^{\tfrac{5}{4}} e^f 
\left( \sin\theta K_i^\psi g {\Ftwo^{i\,\alpha}}{}_\beta \, e^2 + K^\theta_i g {\Ftwo^{i\,\alpha}}{}_\beta \, e^1 \right)
\ .  \\[.4cm]
&\omega^1{}_\alpha = - \frac{5}{4}\frac{X^5(1-x^2) - x^2}{w}  X^{-1}  \partial_\alpha X \, e^1 
+ \frac{1}{2} \ell^{\frac{1}{2}} X^{\tfrac{5}{4}} e^f K_i^\theta g \mathcal{F}^i_{2\,\alpha\beta} \, e^\beta 
\ . \\[.4cm]
&\omega^2{}_\alpha = - \frac{5}{4}\frac{X^5(1-x^2) - x^2}{w}  X^{-1}  \partial_\alpha X \, e^2 
+ \frac{1}{2} \ell^{\frac{1}{2}} X^{\tfrac{5}{4}} e^f \sin\theta K_i^\psi g \mathcal{F}^i_{2\,\alpha\beta} \, e^\beta
\ . \\[.4cm]
&\omega^3{}_\alpha = 
-\ell^{-\frac{1}{2}} X^{-\tfrac{5}{4}} \frac{dA}{dr}  \, e^\alpha 
+ \frac{5}{4} X^{-1} \partial_\alpha X  \, e^3 
\ . \\[.4cm]
&\omega^1{}_2 =  \frac{1}{\sin\theta} \frac{d}{d\psi} \left(K^\theta_i g \A^i\right)  - \cot\theta \, \ell^{-\frac{1}{2}} X^{-\tfrac{5}{4}} e^{-f} \, e^2 
\ . \\[.4cm]
&\omega^1{}_3 = \ell^{-\frac{1}{2}} X^{-\tfrac{5}{4}} \frac{\partial f}{\partial r} \, e^1
\ . \\[.4cm] 
&\omega^2{}_3 = \ell^{-\frac{1}{2}} X^{-\tfrac{5}{4}} \frac{\partial f}{\partial r} \, e^2 
\ . 
\end{align}
\end{subequations}


\bibliography{at}
\bibliographystyle{at}

\end{document}